\documentclass[doublecol,figures,a4paper]{epl2} 

\usepackage{amsmath}
\usepackage{txfonts}

\usepackage[T1]{fontenc}
\usepackage{textcomp}

\usepackage{enumerate}

\usepackage{bm}
\bmdefine{\aVector}{a}
\bmdefine{\AVector}{A}
\bmdefine{\bVector}{b}
\bmdefine{\BVector}{B}
\bmdefine{\cVector}{c}
\bmdefine{\CVector}{C}
\bmdefine{\eVector}{e}
\bmdefine{\EVector}{E}
\bmdefine{\fVector}{f}
\bmdefine{\FVector}{F}
\bmdefine{\gVector}{g}
\bmdefine{\pVector}{p}
\bmdefine{\PVector}{P}
\bmdefine{\qVector}{q}
\bmdefine{\QVector}{Q}
\bmdefine{\rVector}{r}
\bmdefine{\RVector}{R}
\bmdefine{\sVector}{s}
\bmdefine{\uVector}{u}
\bmdefine{\vVector}{v}
\bmdefine{\VVector}{V}
\bmdefine{\muVector}{\mu}
\bmdefine{\OmegaVector}{\Omega}

\title{Molecular simulation of 2-dimensional microphase separation of single-component homopolymers grafted onto a planar substrate}
\shorttitle{microphase separation of single-component grafted homopolymers} 

\author{Y. Norizoe\inst{1} \and H. Jinnai\inst{1,2,3} \and A. Takahara\inst{1,2,3}}
\shortauthor{Yuki Norizoe \etal}

\institute{
	\inst{1} Japan Science and Technology Agency (JST), Exploratory Research for Advanced Technology (ERATO), Takahara Soft Interfaces Project - 
Kyushu University, 744 Motooka, Nishi-ku, Fukuoka 819-0395, Japan  \\
	\inst{2} Institute for Materials Chemistry and Engineering, Kyushu University (IMCE) -
	Kyushu University, 744 Motooka, Nishi-ku, Fukuoka 819-0395, Japan  \\
	\inst{3} International Institute for Carbon-Neutral Energy Research (WPI-I2CNER) -
	Kyushu University, 744 Motooka, Nishi-ku, Fukuoka 819-0395, Japan
}

\abstract{
The structural phase behavior of polymer brushes, single-component linear homopolymers grafted onto a planar substrate, is studied using the molecular Monte Carlo method in 3 dimensions. When simulation parameters of the system are set in regions of macrophase separation of solution for the corresponding non-grafted homopolymers, the grafted polymers also prefer segregation. However, macrophase separation is disallowed due to the spatially-fixed grafting points of the polymers. Such constraints on the grafting are similar to connecting points between blocks of non-grafted diblock copolymers at the microphase separation in the melt state. This results in ``microphase separation'' of the homopolymer brush in the lateral direction of the substrate. Here we extensively search the parameter space and reveal various lateral domain patterns that are similar to those found in diblock copolymer melts at microphase separation.
}

\begin{document}

\maketitle

Polymer chains grafted onto substrates, \textit{i.e.} polymer brushes, have been extensively studied in physics, chemistry, materials science, and other scientific and industrial fields. Polymer brushes are widely utilized for applications such as wetting, adhesion, surface patterning, and colloidal stabilization. Recently, sophisticated polymer brushes, such as brushes composed of diblock copolymers and of binary mixture of homopolymers, have drawn significant attention~\cite{O'Driscoll:2010,Wang:2009}. Polymer brushes on spherical substrates, which are referred to as polymer-grafted colloidal particles, have also attracted wide interest in various scientific and industrial fields~\cite{Griffiths:2011,Norizoe:2005,Norizoe:2012JCP}.

The phase behavior of these complicated brushes is significantly dependent on the molecular architecture, shape of the substrate, ratios between the numbers of polymers of each polymer species, and other characteristics of each system. In the present work, the most basic and simplest polymer brush, \textit{i.e.} single-component linear homopolymers homogeneously grafted onto a planar substrate, is simulated using the molecular Monte Carlo method to determine the universal and structural phase behavior of polymer brushes.

The phase behavior of a solution of single-component non-grafted homopolymers underlies the phase behavior of the present brush. When the chemical and physical parameters of this polymer solution are set in regions within binodal lines of the phase diagram, the non-grafted polymers are segregated into a low-density domain and a high-density domain, which indicates macrophase separation. When the parameters of a layer of the grafted polymers are set in these regions within the binodal lines of the non-grafted polymers, the grafted polymers also prefer segregation. However, due to a topological constraint of the spatially-fixed grafting points of each polymer, macrophase separation of the grafted polymers is disallowed. This constraint has similarity to the topological constraint of connecting points between blocks of non-grafted diblock copolymers at phase separation in the melt state. These non-grafted diblock copolymers exhibit microphase separation and various domain patterns~\cite{Abetz:2005}; therefore, the grafted single-component homopolymers also exhibit phase separation in the lateral direction of the substrate, which results in a variety of 2-dimensional domain patterns of the grafted homopolymers similar to the patterns observed with microphase separation of the non-grafted diblock copolymer melts. These results indicate that the binodal lines of the grafted homopolymer system are changed from those of the non-grafted homopolymers. In the present work, we extensively search the parameter space and reveal the various domain patterns of a 2-dimensional ``microphase separation'' of the grafted homopolymers by measuring the lateral density distribution of the brush.

Early research on single-component linear homopolymers grafted onto planar substrates predicted and confirmed one domain pattern; small circular clusters of the grafted polymers~\cite{Soga:1995,Lai:1992,Tang:1994,Grest:1993,Zhulina:1995,Yeung:1993,Koutsos:1997}. These clusters are distributed on the substrate and are referred to as pinned micelles. The lateral phase separation of the brush is termed microphase separation. However, determination of the spatial arrangement of these pinned micelles was left for future study. Moreover, in this early research, the possibility of the other domain patterns was missed and absent. Here, we reveal a variety of domain patterns and complete the structural phase diagram by extensive simulations.

In the present work, we employ a 3-dimensional solvent-free coarse-grained model~\cite{Drouffe:1991,Daoulas:2010} of single-component linear homopolymers proposed by M{\"{u}}ller and Daoulas~\cite{DoctoralThesis,Norizoe:2010Faraday}. In this model, explicit solvent molecules are integrated out and replaced with an effective non-bonded interaction between solute molecules. This drastically reduces the degrees of freedom of the system and significantly diminishes the computational requirements for simulation of the polymer brush placed in 3 dimensions. The linear molecular architecture of the homopolymers is described by a bead-spring potential:
\begin{equation}
	\label{eq:UsualHarmonicSpringInSingleComponent}
	\frac{ H_{\text{bond}} }{ k_B T } = \sum_{s=1}^{N-1} \frac{3}{2} \frac{ N - 1 }{ R_e^2 } \left\{ \rVector_i (s) - \rVector_i (s+1) \right\}^2 ,
\end{equation}
where $k_B T$ denotes the thermal energy, $\rVector_i (s)$ represents the coordinate of the $s$-th coarse-grained segment of the $i$-th polymer, and $N$ is the number of segments per polymer. $R_e$, which denotes the root mean square of the end to end distance of an ideal chain with the same molecular architecture, is chosen as the unit length. $H_{\text{bond}}$ represents the bonded interaction potential between the segments. On the other hand, the free energy of non-bonded interactions in the solvent-free model is given as a functional of the local segment density. A third-order expansion of the non-bonded interaction free energy in a form of powers of the local segment density is employed:
\begin{equation}
	\label{eq:SingleComponentNonidealFreeEnergy}
	\frac{ H_{\text{non-bonded}} }{k_B T} := \int_V dV \left( -\frac{1}{2}v \left( \rho_s ( \rVector ) \right)^2 + \frac{1}{3}w \left( \rho_s ( \rVector ) \right)^3 \right),
\end{equation}
where $\rho_s ( \rVector )$ is the local volumetric number density of the segments at the spatial position $\rVector$. The positive constants, $v$ and $w$, correspond to the attractive and repulsive interaction strengths among the segments, respectively. In the following calculation, dimensionless physical quantities are utilized: the local polymer density $\rho_p' ( \rVector ) = \rho_s ( \rVector ) R_e^3 / N$, the parameter $w' = w N^3 / {R_e}^6$, and the parameter $v' = v N^2 / R_e^3$. These dimensionless quantities reduce eq.~\eqref{eq:SingleComponentNonidealFreeEnergy} to:
\begin{equation}
\label{eq:SingleComponentNonidealFreeEnergyDimensionless}
		\frac{ H_{\text{non-bonded}} }{k_B T} = \int_V \frac{dV}{{R_e}^3} \left( -\frac{1}{2}v' \left( \rho_p' ( \rVector ) \right)^2 + \frac{1}{3}w' \left(\rho_p' ( \rVector ) \right)^3 \right).
\end{equation}

\begin{figure}[!tb]
	\centering
	\includegraphics[clip]{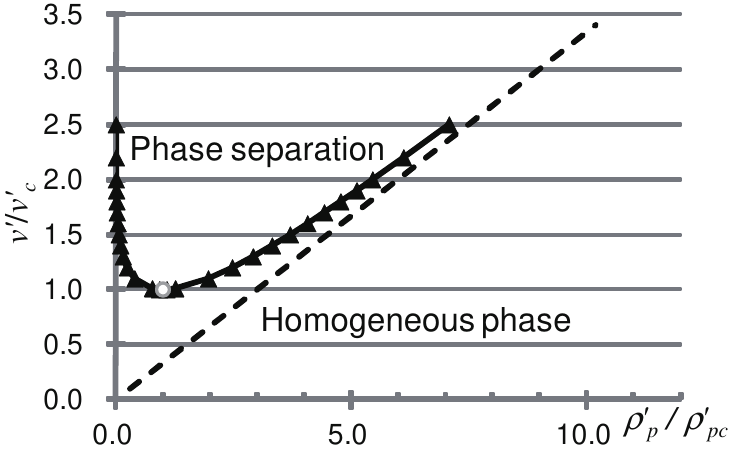}
	\caption{Binodal lines of a system composed of non-grafted polymers at fixed $w' = 0.0001$, constructed based on the mean-field theory. The critical point, $( \rho_{pc}' , v'_c )$, is denoted by a circle. The broken line represents the asymptotic behavior of the binodal line at high polymer density. The validity of this phase diagram is confirmed by molecular Monte Carlo simulation.}
	\label{fig:BinodalLineNongrafted}
\end{figure}

First, we discuss the phase behavior of a solution of the non-grafted polymers modeled using the solvent-free system. The dimensionless average volumetric polymer density is denoted by $\rho_p' = n_p R_e^3 / V$, where $n_p$ and $V$ are the number of polymers and the system volume, respectively. At extremely large $v'$ and finite $w'$, the polymers aggregate, which results in one dense droplet floating in a dilute gas, which indicates macrophase separation of the polymers. In contrast, at finite $v'$ and extremely high $w'$, the system is a homogeneous phase. The binodal lines in the $\rho_p' v'$-plane at fixed $w'$ are calculated based on mean-field theory~\cite{DoctoralThesis,Doi:IntroductionToPolymerPhysics}. The critical point of this phase diagram is denoted by $\rho_p' = \rho_{pc}'$ and $v' = v'_c$~\cite{DoctoralThesis}:
\begin{equation}
	\label{eq:SingleComponentCriticalPointDimensionless}
	\left( \rho_{pc}' = \frac{1}{\sqrt{ 2w' }} \, , \, v'_c = 2 \sqrt{2w'} \, \right).
\end{equation}
At high average polymer density, the fluctuation of the polymer density becomes small, so that the mean-field theory provides an accurate prediction of the binodal lines. As an example, the result of the mean-field calculation at fixed $w' = 0.0001$ is shown in Fig.~\ref{fig:BinodalLineNongrafted}. At this value of $w'$, $\rho_{pc}' = 70.71$. This high $\rho_{pc}'$~\cite{DoctoralThesis} provides a good prediction of the binodal line. This phase diagram shows that the parameter $v'$ corresponds to the inverse temperature, which indicates that the quality of the implicit solvents at fixed $w'$ decreases with $v'$. Based on this phase diagram of the polymer solution, the polymer brush is simulated and the structural phase diagram is constructed at $w' = 0.0001$.

In this simulation, $\rho_p' ( \rVector )$ is calculated \textit{via} a collocation lattice. For this purpose, the system box is partitioned into a cubic lattice, denoted by $\{ \aVector \}$, with $\varDelta L$ of the grid spacing. According to related particle-to-mesh methods in electrostatics~\cite{Eastwood:1980,Deserno:1998,Mueller:2008JCP}, the densities at each grid point, $\aVector$, are determined as:
\begin{equation}
\label{eq:PPPMPolymerDensityOnGrid}
	\rho_p' ( \aVector ) = \frac{ R_e^3 }{ N } \sum_{i=1}^{n_p} \sum_{s=1}^N \Phi \left( \rVector_i (s), \aVector \right).
\end{equation}
The function in this equation, $\Phi \left( \rVector, \aVector \right)$, allocates the particles to grid points, according to a linear assignment function,
\begin{gather}
\label{eq:PPPMLinearAssignmentFunction}
	\Phi \left( \rVector, \aVector \right) = \frac{1}{\varDelta L^3} \prod_{\alpha = x,y,z} g \left( d_\alpha \right),  \\
	g \left( d_\alpha \right) = \begin{cases}
									 1 - |d_\alpha| / \varDelta L &\text{for} \quad |d_\alpha| < \varDelta L  \\
									 0  & \text{otherwise}
	\end{cases}
\end{gather}
where $d_\alpha = r_\alpha - a_\alpha$ denotes the distance between the grid point, $\aVector$, and the segment position, $\rVector$, along the Cartesian direction, $\alpha$. The grid spacing, $\varDelta L$, defines the range of the non-bonded interaction. Utilizing this grid-based density, the non-bonded interaction of eq.~\eqref{eq:SingleComponentNonidealFreeEnergyDimensionless} from $\rho_p' ( \aVector )$ is calculated by substituting the summation over the lattice nodes $\sum_\aVector \varDelta L^3$ for the integral $\int_V dV$.

In other words, in the present calculation of the non-bonded interaction \textit{via} the collocation grid, the presence of each particle is divided into the eight grid points surrounding this particle~\cite{DoctoralThesis}. We assume that the coordinates of this particle, denoted by $( r_x, r_y, r_z )$, satisfies a relation $j_\alpha \le r_\alpha / \varDelta L < j_\alpha + 1$, where $( j_x, j_y, j_z )$ denotes the index of the grid points. This means that the particle is surrounded by the eight grid points $( j_x, j_y, j_z )$, $( j_x, j_y, j_z + 1 )$, $( j_x, j_y + 1, j_z ), \dots, ( j_x + 1, j_y + 1, j_z + 1 )$. The contribution from this particle to the dimensionless average polymer density in the cubic cell of these eight grids is $(1 / N \varDelta L^3 ) R_e^3$. In the present simulation, this contribution is linearly interpolated among these eight grid points:
\begin{equation}
	\label{eq:DividedPolymerDensityGrids}
	\begin{split}
		&\phi '_{ j_x, j_y, j_z } = \frac{
				\left( u^{(x)}_{ j_x + 1 } - r_x \right) \left( u^{(y)}_{ j_y + 1 } - r_y \right) \left( u^{(z)}_{ j_z + 1 } - r_z \right)
			}
			{
				\varDelta L^3
			}
			\frac{ R_e^3 }{ N \varDelta L^3 } \, ,  \\
		&\phi '_{ j_x, j_y, j_z + 1 } = \frac{
				\left( u^{(x)}_{ j_x + 1 } - r_x \right) \left( u^{(y)}_{ j_y + 1 } - r_y \right) \left( r_z - u^{(z)}_{ j_z } \right)
			}
			{
				\varDelta L^3
			}
			\frac{ R_e^3 }{ N \varDelta L^3 } \, ,  \\
		&\phi '_{ j_x, j_y + 1, j_z } = \frac{
				\left( u^{(x)}_{ j_x + 1 } - r_x \right) \left( r_y - u^{(y)}_{ j_y } \right) \left( u^{(z)}_{ j_z + 1 } - r_z \right)
			}
			{
				\varDelta L^3
			}
			\frac{ R_e^3 }{ N \varDelta L^3 } \, ,  \\
		&\\
		&\dots\\
		&\\
		&\phi '_{ j_x + 1, j_y + 1, j_z + 1 } = \frac{
				\left( r_x - u^{(x)}_{ j_x } \right) \left( r_y - u^{(y)}_{ j_y } \right) \left( r_z - u^{(z)}_{ j_z } \right)
			}
			{
				\varDelta L^3
			}
			\frac{ R_e^3 }{ N \varDelta L^3 } \, ,
	\end{split}
\end{equation}
where $\phi '$ denotes the interpolated dimensionless polymer density at each surrounding grid point, and $u^{(\alpha)}_{ \nu_\alpha } = \nu_\alpha \varDelta L$ is the $\alpha$-coordinate of the grid point specified by the index $\nu_\alpha$. This $\phi '$ corresopnds to the linear assignment function, $\Phi$, of eq.~\eqref{eq:PPPMLinearAssignmentFunction}. We can obtain the dimensionless total polymer density at each grid point by summing $\phi '$ over all the particles adjacent to the grid point. This results in eq.~\eqref{eq:PPPMPolymerDensityOnGrid}. The obtained dimensionless total polymer density at each grid point is substituted for the non-bonded interaction reduced \textit{via} the collocation grid:
\begin{equation}
\label{eq:SingleComponentNonidealFreeEnergyDimensionlessGrids}
	\frac{ H_{\text{non-bonded}} }{k_B T} = \sum_\aVector \left\{ \frac{ \varDelta L^3 }{{R_e}^3} \left( -\frac{1}{2}v' \left( \rho_p' ( \aVector ) \right)^2 + \frac{1}{3}w' \left(\rho_p' ( \aVector ) \right)^3 \right) \right\}.
\end{equation}
This calculation of the non-bonded interaction \textit{via} the collocation grid is computationally advantageous, because a segment in this simulation interacts with many neighbors.

Monte Carlo simulations are performed with the canonical ensemble in 3 dimensions \textit{via} the standard Metropolis algorithm~\cite{ComputerSimulationOfLiquids,Frenkel:UnderstandingMolecularSimulation2002} at the fixed $w' = 0.0001$. This value of $w'$ is the same as that for the phase diagram in Fig.~\ref{fig:BinodalLineNongrafted}. $k_B T$ is chosen as the unit energy. $(L_x, L_y, L_z)$ denotes the size of the rectangular parallelepiped system box and is fixed at $( L_x = 12.0 R_e, L_y = 12.0 R_e, L_z = 6.0 R_e )$. A periodic boundary condition is applied to the system. The system box is placed in spatial regions of $0 \le \alpha < L_\alpha$. $N = 32$ and $\varDelta L = (1/6) R_e$ are fixed. The Mersenne Twister algorithm is selected as a random number generator for the simulations~\cite{MersenneTwister1,MersenneTwister2,MersenneTwister3}. In one simulation step, a particle is picked at random and given a uniform random trial displacement within a cube of edge length $2 \varDelta L$. A Monte Carlo step (MCS) is defined as $n_p N$ trial moves, during which each particle is selected for the trial displacement once on average. In simulation with these simulation parameters, the mean square end-to-end distance of an isolated non-grafted polymer becomes $1.00 R_e^2$ at $v'_\theta / v'_c = v' / v'_c = 0.1$. This indicates that this $v'_\theta$ corresponds to the inverse theta temperature, and that the implicit solvents in regions of $v' > v'_\theta$ correspond to poor solvents.

One end segment of each polymer is grafted onto and fixed at a homogeneous square lattice on a thin hard planar wall placed at $z = 0$. This means that the polymers are homogeneously grafted onto the $xy$-plane of the system box. This substrate disallows the polymers to pass through. No other interactions, such as coulombic and frictional interactions, act between the substrate and the polymers. The other segments of the grafted polymers are built in $z > 0$ and form a polymer layer in an interval of $0 < z \lesssim 1 R_e$. The average volumetric polymer density in this region of the layer, defined as $\rho_p = n_p / ( L_x L_y \times 1 R_e )$, corresponds to the average volumetric polymer density in the system of the non-grafted polymers. Simulations are performed for various value sets of the parameters $( \rho_p' = \rho_p R_e^3, v' )$ of the brush, and the structural phase diagram is constructed in the $\rho_p' v'$-plane. We have confirmed that this phase diagram for the brush is not significantly changed when the graft points of the polymers are randomly distributed over the substrate using uniform random numbers, which results in an inhomogeneous spatial distribution of the graft points.

In the initial state, the conformation of each polymer is randomly arranged according to a Gaussian distribution, which results in random coils~\cite{Kawakatsu:StatisticalPhysicsOfPolymersAnIntroduction} with $\approx \! \! 1 R_e$ of the root mean square end-to-end distance.

To characterize the domain patterns, the local lateral density of the grafted polymers is defined as:
\begin{equation}
\label{eq:ArealPolymerDensity}
	\rho_p^{(\textit{A})} ( x, y ) = \int_0^{L_z} dz \, \frac{ \rho_s ( x, y, z ) }{ N },
\end{equation}
which has a dimension of areal density. The average areal density of the polymer brush is denoted by $\rho_p^{(\textit{A})} = n_p / L_x L_y$. Due to the constraint of the graft, except at extremely high $v'$, the periodicity of the domain patterns is $\sim \! \! R_e$, because the bonded interaction energy and conformational entropy of the polymers disallow the polymers to stretch $\gg \! \! R_e$ or to shrink $\ll \! \! R_e$.

\begin{figure}[!tb]
	\centering
	\includegraphics[clip]{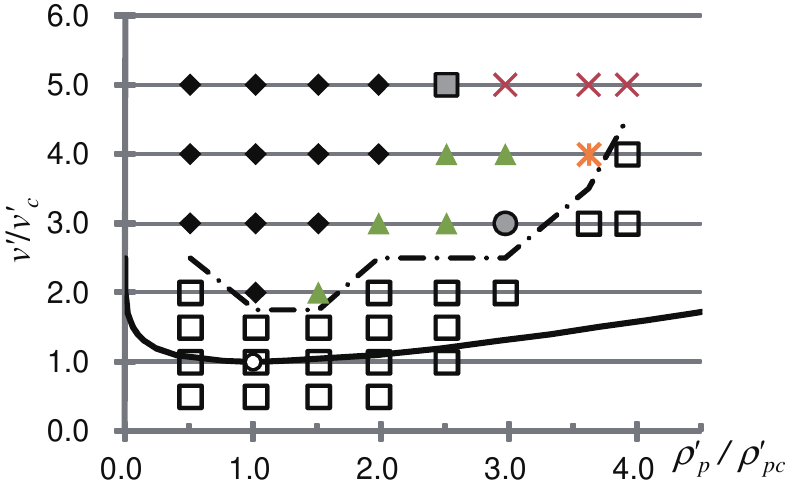}
	\caption{Phase diagram of the polymer brush, traced on the phase diagram of the non-grafted polymers given in Fig.~\ref{fig:BinodalLineNongrafted}. A chain line denotes the binodal line of the brush. White squares represent the homogeneous phase of the brush. Diamonds, triangles, and crosses represent the hexagonal, lamella, and inverse hexagonal structures, respectively. These three structures are simultaneously observed in the system at the point of the grey square. The grey circle denotes the inverse nematic structure and the asterisk is a structure possibly similar to, but different from, the lamellae.}
	\label{fig:PhaseDiagramBrushW00001}
\end{figure}

\begin{figure}[!tb]
	\centering
	\includegraphics[clip]{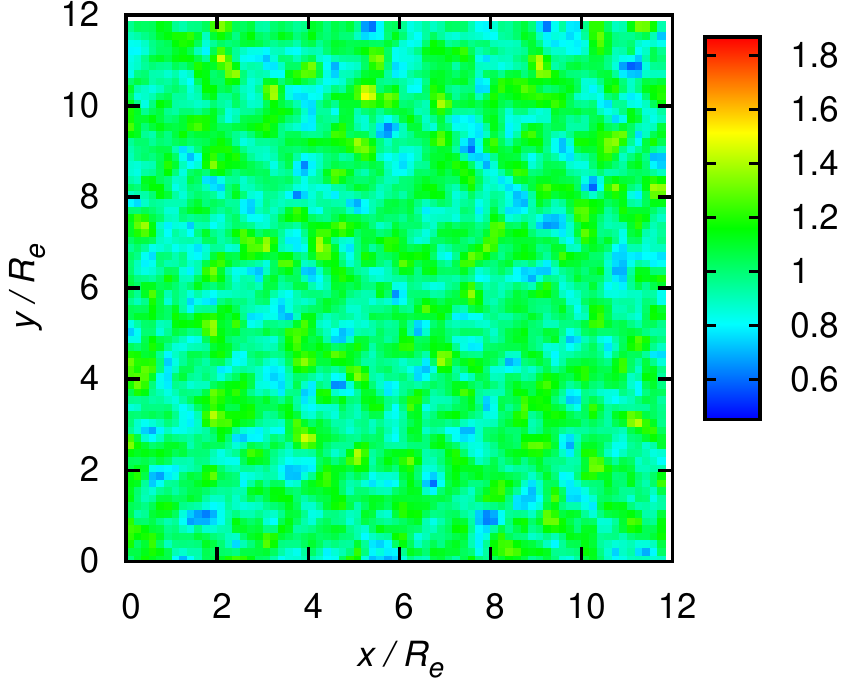}
	\caption{Lateral polymer density distribution, $\rho_p^{(\textit{A})} ( x, y ) / \rho_p^{(\textit{A})}$, at $\rho_p' / \rho_{pc}' = 1.02$, $v' / v'_c = 0.5$, and $1 \times 10^6$ MCS. $n_p = 10404$.}
	\label{fig:BrushIniSqLatN32Lx120Ly120Lz60RRc102VVc05_001000000MCS-SpatialSegmentDistributionForColumnsZ}
\end{figure}

Here, we discuss the simulation results for the brush. The phase diagram of the brush, which is traced on the phase diagram of the non-grafted polymers given in Fig.~\ref{fig:BinodalLineNongrafted}, is shown in Fig.~\ref{fig:PhaseDiagramBrushW00001}. In regions of $v' / v'_c$ below a chain line drawn in this phase diagram of the $\rho_p' v'$-plane, the homogeneous phase of the brush is maintained until the simulation halts at $\sim \! \! 1 \times 10^6$ MCS. As an example of this homogeneous structure, $\rho_p^{(\textit{A})} ( x, y ) / \rho_p^{(\textit{A})}$ at $\rho_p' / \rho_{pc}' = 1.02$ and $v' / v'_c = 0.5$ is presented in Fig.~\ref{fig:BrushIniSqLatN32Lx120Ly120Lz60RRc102VVc05_001000000MCS-SpatialSegmentDistributionForColumnsZ}. On the other hand, in regions of $v' / v'_c$ above the chain line, $\rho_p^{(\textit{A})} ( x, y )$ indicates the 2-dimensional microphase separation of the grafted polymers. The domains are spontaneously created at less than $\sim \! \! 4 \times 10^5$ MCS and remain until the simulation halts at $\sim \! \! 1 \times 10^6$ MCS, without significant change of the positions, shapes, and the sizes. This indicates that the chain line denotes the binodal line of the polymer brush. This binodal line is shifted from that of the non-grafted polymers toward large $v' / v'_c$. However, these two binodal lines are still qualitatively consistent, as expected. Next, we discuss the domain patterns for the 2-dimensional microphase separation of the polymers.

\begin{figure*}[!tb]
	\centering
	\includegraphics[clip]{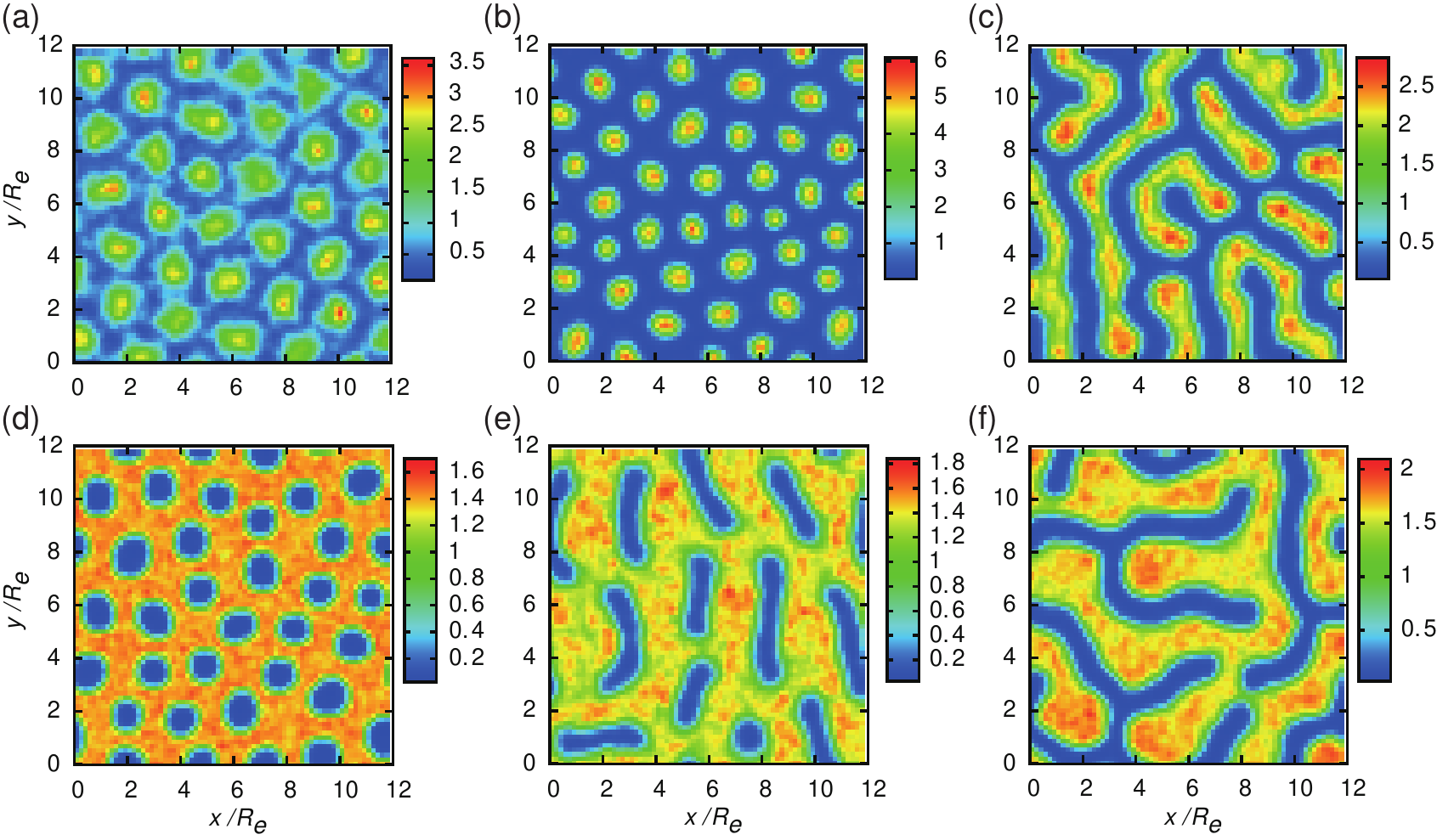}
	\caption{$\rho_p^{(\textit{A})} ( x, y ) / \rho_p^{(\textit{A})}$ at $1 \times 10^6 $ MCS. Results at (a): $( \rho_p' / \rho_{pc}' = 1.02, v' / v'_c = 2.0 )$, (b): $( 1.02, 3.0 )$, (c): $( 1.98, 3.0 )$, (d): $( 3.92, 5.0 )$, (e): $( 2.97, 3.0 )$, and (f): $( 3.62, 4.0 )$.}
	\label{fig:Brush2DVariousStructures}
\end{figure*}

At $\rho_p' / \rho_{pc}' = 1.02$, circular clusters assembled in the hexagonal structure are observed at $v' / v'_c = 2.0$, as shown in Fig.~\ref{fig:Brush2DVariousStructures}(a). In this figure, $\approx \! \! 40$ circular clusters are found. With this value of $\rho_p' / \rho_{pc}'$, the number of polymers becomes $n_p = 10404$ in the present simulation system, which indicates that each circular cluster is composed of an order of 100 polymers. This hexagonal structure is also observed in the region of $0.509 \lesssim \rho_p' / \rho_{pc}' \lesssim 1.98$ and $2.0 \lesssim v' / v'_c \lesssim 5.0$, which is represented by diamonds in Fig.~\ref{fig:PhaseDiagramBrushW00001}. At high $v' / v'_c$ in this region, rearrangement of the clusters requires a long time, due to the high interfacial energy barrier. This results in long-lived defects in the hexagonal structure and inhomogeneity of the cluster size, as shown in Fig.~\ref{fig:Brush2DVariousStructures}(b), as an example, at $\rho_p' / \rho_{pc}' = 1.02$ and $v' / v'_c = 3.0$.

When $\rho_p' / \rho_{pc}'$ is raised from the region of the hexagonal structure, the clusters increase in size and merge with each other. However, due to the constraint of the graft, the periodicity of the domain patterns is constrained to $\sim \! \! R_e$. Therefore, large circular clusters with sizes $\gg \! \! R_e$ arranged in the hexagonal structure are not created in the system. Therefore, when $\rho_p' / \rho_{pc}'$ rises, the hexagonal structure changes to another structure, which is denoted by triangles in Fig.~\ref{fig:PhaseDiagramBrushW00001}. Figure~\ref{fig:Brush2DVariousStructures}(c) gives an example of this structure, \textit{i.e.} 2-dimensional lamella structure, which is sampled at $\rho_p' / \rho_{pc}' = 1.98$ and $v' / v'_c = 3.0$.

At $v' / v'_c = 5.0$, the inverse hexagonal structure is determined in the interval of $2.97 \le \rho_p' / \rho_{pc}' \le 3.92$. In this structure, circular voids are created in the dense polymer layer of the brush, as shown in Fig.~\ref{fig:Brush2DVariousStructures}(d). However, at $\rho_p' / \rho_{pc}' = 2.51$, the hexagonal, lamella, and inverse hexagonal structures simultaneously appear in the system.

At $\rho_p' / \rho_{pc}' = 2.97$ and $v' / v'_c = 3.0$, sausage-like voids are formed in the dense polymer layer, as shown in Fig.~\ref{fig:Brush2DVariousStructures}(e). These anisotropic shaped voids lie in the same direction locally, which is similar to the nematic phase of liquid crystals.

At $\rho_p' / \rho_{pc}' = 3.62$ and $v' / v'_c = 4.0$, a structure similar to the lamellae is observed. However, in this structure, low-density layers of the lamellae compose junction points similar to tripods, as presented in Fig.~\ref{fig:Brush2DVariousStructures}(f). Thus, each junction point is composed of and homogeneously connected with the 3 low-density layers, which indicates that a structure different from the lamellae is evident at this state point.

Finally, the angular distribution of the end-to-end vectors of the grafted polymers is examined. When the graft point of a polymer is located in regions of low $\rho_p^{(\textit{A})} ( x, y )$ in 2-dimensional microphase separation, this polymer is laterally stretched toward regions of high $\rho_p^{(\textit{A})} ( x, y )$. To confirm this effect, the latitude $\theta$, of the end-to-end vector, \textit{i.e.} a vector from the grafted segment to the free end of each polymer, is measured. In this simulation, the latitudinal distribution defined as,
\begin{equation}
\label{eq:DefinitionOfDistributionOfLatitude}
	f( \theta ) :=
	\frac{ p_{\text{latitude}} \left( \theta \right) \, d\theta }{ \frac{ 1 }{ 2 } \left\{ \cos \theta - \cos \left( \theta + d \theta \right) \right\} },
\end{equation}
is calculated, where $p_{\text{latitude}} (\theta)$ denotes the probability density of $\theta$. When the vectors are homogeneously distributed over all directions for the latitude and longitude, $f( \theta )$ becomes homogeneous, \textit{i.e.} $f( \theta ) = 1$ over the entire interval of $0 \le \theta \le \pi$. Therefore, the inhomogeneity of the distribution in the latitudinal direction is quickly evident in $f( \theta )$.

\begin{figure}[!htbp]
	\centering
	\includegraphics[clip]{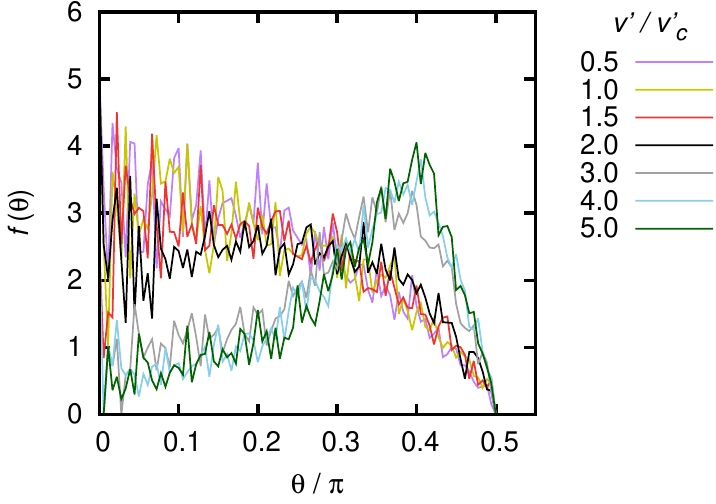}
	\caption{$f( \theta )$ at $\rho_p' / \rho_{pc}' = 0.509$ and $1 \times 10^6$ MCS for various values of $v' / v_c'$. $f( \theta ) = 0$ for $\theta / \pi \ge 0.5$, due to the hard substrate placed on the $xy$-plane.}
	\label{fig:BrushIniSqLatN32Lx120Ly120Lz60RRc0509_001000000MCS-LatitudeProbDensity}
\end{figure}

Figure~\ref{fig:BrushIniSqLatN32Lx120Ly120Lz60RRc0509_001000000MCS-LatitudeProbDensity} shows $f( \theta )$ at $\rho_p' / \rho_{pc}' = 0.509$. At $v' / v_c' \le 2.0$, \textit{i.e.} in the homogeneous phase, a peak is found at $\theta \approx 0$. This is similar to densely grafted polymers in the homogeneous phase in good solvents, in which the grafted polymers stretch away from the substrate due to the excluded volume interaction between adjacent polymers~\cite{Milner:1991,Koutsos:1997}. In contrast, the regions of 2-dimensional phase separation, \textit{i.e.} $v' / v_c' \ge 3.0$, reveal another sharp peak at $\theta / \pi \approx 0.4$, in which the peak at $\theta \approx 0$ is covered. This indicates that the grafted polymers tend to be stretched laterally, as expected. This result corroborates the location of the binodal line given in Fig.~\ref{fig:PhaseDiagramBrushW00001}.

In conclusion, single-component linear homopolymers homogeneously grafted onto a planar substrate have been simulated in 3 dimensions using the molecular Monte Carlo method. The 2-dimensional microphase separation of grafted polymers in the lateral direction of the substrate has been examined and a variety of lateral domain patterns have been revealed. This result is similar to that for the 3-dimensional microphase separation of non-grafted diblock copolymer melts. The structural phase diagram of a homopolymer brush has been constructed, the binodal line of which is qualitatively consistent with that for the non-grafted homopolymer system.

Both non-grafted block copolymer melts and grafted homopolymers exhibit microphase separation and various domain patterns. A block copolymer is composed of chemically immiscible blocks that are separated into microstructures. However, such phenomena are also observed for the grafted homopolymers, despite the absence of chemical immiscibility along each polymer chain and distinct molecular architectures. This indicates that the topological constraints, \textit{i.e.} the grafting points of homopolymer brushes and the connecting points between the blocks in each block copolymer, are the origin of microphase separation and the various domain patterns. This also demonstrates that both the intramolecular topological constraints, \textit{i.e.} the connecting points between the blocks, and extramolecular topological constraints, \textit{i.e.} the grafting points, yield these phenomena. This result is a universal characteristic of arbitrary polymers.

\acknowledgments
The authors wish to thank Prof. Marcus M{\"u}ller, Dr Kostas Ch. Daoulas, and Dr Taiki Hoshino for helpful suggestions and discussions. H. Jinnai gratefully acknowledges the financial support received through a Grant-in-Aid (No. 24310092) from the Ministry of Education, Culture, Sports, Science, and Technology (MEXT) of Japan.


\begin{thebibliography}{10}
\expandafter\ifx\csname url\endcsname\relax\def\url#1{\texttt{#1}}\fi

\bibitem{O'Driscoll:2010}
\Name{O'Driscoll B. M.~D., Griffiths G.~H., Matsen M.~W., Perrier S., Ladmiral
  V. \and Hamley I.~W.} \REVIEW{Macromolecules }{43}{2010}{8177}.

\bibitem{Wang:2009}
\Name{Wang J. \and M{\"{u}}ller M.} \REVIEW{J. Phys. Chem. B
  }{113}{2009}{11384}.

\bibitem{Griffiths:2011}
\Name{Griffiths G.~H., Vorselaars B. \and Matsen M.~W.} \REVIEW{Macromolecules
  }{44}{2011}{3649}.

\bibitem{Norizoe:2005}
\Name{Norizoe Y. \and Kawakatsu T.} \REVIEW{{Europhys. Lett.} }{72}{2005}{583}.

\bibitem{Norizoe:2012JCP}
\Name{Norizoe Y. \and Kawakatsu T.} \REVIEW{J. Chem. Phys.
  }{137}{2012}{024904}.

\bibitem{Abetz:2005}
\Name{Abetz V. \and Simon P. F.~W.} \Book{Phase behaviour and morphologies of
  block copolymers} in \Book{Block Copolymers I}, edited by \Name{Abetz V.}
  Vol. 189 of \emph{Adv. Polym. Sci.} (Springer) 2005 pp. 125--212.

\bibitem{Soga:1995}
\Name{Soga K.~G., Guo H. \and Zuckermann M.~J.} \REVIEW{Europhys. Lett.
  }{29}{1995}{531}.

\bibitem{Lai:1992}
\Name{Lai P.-Y. \and Binder K.} \REVIEW{J. Chem. Phys. }{97}{1992}{586}.

\bibitem{Tang:1994}
\Name{Tang H. \and Szleifer I.} \REVIEW{Europhys. Lett. }{28}{1994}{19}.

\bibitem{Grest:1993}
\Name{Grest G.~S. \and Murat M.} \REVIEW{Macromolecules }{26}{1993}{3108}.

\bibitem{Zhulina:1995}
\Name{Zhulina E.~B., Birshtein T.~M., Priamitsyn V.~A. \and Klushin L.~I.}
  \REVIEW{Macromolecules }{28}{1995}{8612}.

\bibitem{Yeung:1993}
\Name{Yeung C., Balazs A.~C. \and Jasnow D.} \REVIEW{Macromolecules
  }{26}{1993}{1914}.

\bibitem{Koutsos:1997}
\Name{Koutsos V., van~der Vegte E.~W., Pelletier E., Stamouli A. \and
  Hadziioannou G.} \REVIEW{Macromolecules }{30}{1997}{4719}.

\bibitem{Drouffe:1991}
\Name{Drouffe J.~M., Maggs A.~C. \and Leibler S.} \REVIEW{Science
  }{254}{1991}{1353}.

\bibitem{Daoulas:2010}
\Name{Daoulas K.~C. \and M{\"{u}}ller M.} \REVIEW{Adv. Polym. Sci.
  }{224}{2010}{197}.

\bibitem{DoctoralThesis}
\Name{Norizoe Y.} \Book{Measuring the free energy of self-assembling systems in
  computer simulation} Ph.D. thesis Institute for Theoretical Physics,
  University of G{\"o}ttingen, G{\"o}ttingen, Germany (January 2010).
\newline\url{http://webdoc.sub.gwdg.de/diss/2010/norizoe/}

\bibitem{Norizoe:2010Faraday}
\Name{Norizoe Y., Daoulas K.~C. \and M{\"{u}}ller M.} \REVIEW{Faraday Discuss.
  }{144}{2010}{369}.

\bibitem{Doi:IntroductionToPolymerPhysics}
\Name{Doi M.} \Book{Introduction to Polymer Physics} (Oxford University Press,
  Oxford) 1996.

\bibitem{Eastwood:1980}
\Name{Eastwood J.~W., Hockney R.~W. \and Lawrence D.~N.} \REVIEW{Computer
  Physics Communications }{19}{1980}{215}.

\bibitem{Deserno:1998}
\Name{Deserno M. \and Holm C.} \REVIEW{J. Chem. Phys. }{109}{1998}{7678}.

\bibitem{Mueller:2008JCP}
\Name{M{\"{u}}ller M. \and Daoulas K.~C.} \REVIEW{J. Chem. Phys.
  }{128}{2008}{024903}.

\bibitem{ComputerSimulationOfLiquids}
\Name{Allen M.~P. \and Tildesley D.~J.} \Book{Computer Simulation of Liquids}
  (Oxford University Press, Oxford) 1989.

\bibitem{Frenkel:UnderstandingMolecularSimulation2002}
\Name{Frenkel D. \and Smit B.} \Book{Understanding molecular simulation: from
  algorithms to applications} (Academic Press, London) 2002.

\bibitem{MersenneTwister1}
\Name{Matsumoto M. \and Nishimura T.} \REVIEW{ACM Trans. Model. Comput. Simul.
  }{8}{1998}{3}.

\bibitem{MersenneTwister2}
\Name{Matsumoto M. \and Kurita Y.} \REVIEW{ACM Trans. Model. Comput. Simul.
  }{2}{1992}{179}.

\bibitem{MersenneTwister3}
\Name{Matsumoto M. \and Kurita Y.} \REVIEW{ACM Trans. Model. Comput. Simul.
  }{4}{1994}{254}.

\bibitem{Kawakatsu:StatisticalPhysicsOfPolymersAnIntroduction}
\Name{Kawakatsu T.} \Book{Statistical Physics of Polymers: An Introduction}
  (Springer) 2004.

\bibitem{Milner:1991}
\Name{Milner S.~T.} \REVIEW{Science }{251}{1991}{905}.

\end{thebibliography}

\end{document}